\documentclass[aip,amsmath,amssymb]{revtex4-1}

\usepackage{amssymb,amsmath}
\usepackage{graphicx} 
\usepackage{epstopdf}
\usepackage{bm} 
\begin{document}

\title{Complex Network Approach to Fractional Time Series}
\author{Pouya Manshour}
\affiliation{Physics Department, Persian Gulf University, Bushehr 75169, Iran}

\begin{abstract}
In order to extract correlation information inherited in stochastic time series, the visibility graph algorithm has been recently proposed, by which a time series can be mapped onto a complex network. We demonstrate that the visibility algorithm is not an appropriate one to study the correlation aspects of a time series. We then employ the \emph{horizontal} visibility algorithm, as a much simpler one, to map fractional processes onto complex networks. The degree distributions are shown to have parabolic exponential forms with Hurst dependent fitting parameter. Further, we take into account other topological properties such as maximum eigenvalue of the adjacency matrix and the degree assortativity, and show that such topological quantities can also be used to predict the Hurst exponent, with an exception for anti-persistent fractional Gaussian noises. To solve this problem, we take into account the Spearman correlation coefficient between nodes' degrees and their corresponding data values in the original time series.
\end{abstract}


\maketitle

\begin{quotation}
To map a time series into a complex network with the aim of extracting useful information inherited in the time series by studying various topological features of the resulting network, has been subject of intense studies in recent years. The visibility graph algorithm, among others, has proven itself to be a powerful tool to study correlation structures of chaotic as well as stochastic time series. In this work, we investigate correlation aspects of fractional time series by studying various topological characteristics of the associated horizontal visibility graph. Such topological quantities are able to predict the Hurst exponent for fractional processes, except for anti-persistent fractional Gaussian noises. We then solve this problem by proposing a quantity, which takes into account the correlation between nodes' degrees in the graph and values of the data points in the original time series.
\end{quotation}
\section{Introduction}
To investigate time series as the output of many natural systems has been one of the most challenging fields of study for many years. Commonly, our access to the dynamical origin of natural systems is limited, thus the time behavior of their response functions may be our only asset. Time series analysis, as a powerful tool, try to extract possible underlying forces and structures that construct the observed data or to fit a model for forecasting or control. This formalism can apply to real data from physics, biology, economy, medicine, engineering, and among others \cite{Kantz1997,Shumway2000}. In order to model observed time series in real world, many stochastic models have been proposed \cite{Karlin1998,Lemons2002}. In $1940$, Kolmogorov originally introduced \cite{Kolmogorov1940} a stochastic process in order to model turbulent flows, which was after named ``fractional Brownian motion" (fBm) in the seminal paper by Mandelbrot and Van Ness \cite{Mandelbrot1968}. The fractional Brownian motion, a generalization of the more well-known Brownian motion, has been one of the most studied stochastic processes used in a variety of fields, including physics, probability, statistics, hydrology, economy, biology, and many others \cite{Mandelbrot2002,Embrechts2002,Karag2004,Karmeshu2004,Robinson2003,Varotsos2006,Alvarez2006}. The fBm is a self-similar Gaussian process with stationary increments (fractional Gaussian noise-fGn) and possesses long-memory which depends on a parameter $H\in(0,1)$ called the Hurst index \cite{Hurst1951}. The case $H=1/2$ corresponds to the ordinary Brownian motion in which successive increments are statistically independent of one another and thus there is no correlation. For $H>1/2$ the increments of the process are positively correlated (persistent) and for $H<1/2$ consecutive increments are more likely to have opposite signs (anti-persistent). Estimating the Hurst exponent for a process provides a measure of whether the data is a pure white noise random process or has underlying trend. Thus, the aim of numerous studies is to uncover correlation information of an empirical time series, by calculating its corresponding Hurst index. However, calculating the Hurst exponent of a time series is a tricky task, and a variety of techniques have been proposed \cite{Hurst1951,Holsch1988,Peng1994,Vandewalle1998}. However, each of these methods has specific advantages and disadvantages, and the accuracy of the estimation can also be questioned, therefore, the search for alternatives is obvious.

In recent years, the complex network theory has attracted much attention, with the aim of studying various possible information of a complex system by using some concepts borrowed from graph theory as well as statistical physics \cite{Barabasi2002,Barrat2008,Pastor2007,Murray1993,Montakhab2012}. More recently, a graph-theoretical approach in time series analysis has been developed, and the network-based theories have been applied in many disciplines such as biology, sociology, physics, climatology, and neurosciences \cite{Zhang2006,Lacasa2008,Xu2008,Shirazi2009,marwan2009,Donner2010,Donner2011,Manshour2015}. In this approach, a time series is mapped into a (complex) graph, and the characteristics of the time series are believed to be inherited in the resulting network, which can be analyzed from a complex network perspective. One of the most studied of these maps has been proposed by Lacasa \textit{et al.} in $2008$. They introduced an algorithm \cite{Lacasa2008}, called Visibility Graph (VG), which maps a time series into a graph based on the ability of the data points to see each other and it is defined as follows: Let $x_i$ be a time series of size $N$ ($i=1,2, ... N$). The algorithm assigns each datum of the series to a node in the VG. Accordingly, a series of size $N$ map to a graph with $N$ nodes. Two nodes $i$ and $j$ in the graph are connected if one can draw a (straight) line in the time series joining $x_i$ and $x_j$ that does not intersect any intermediate data height, i.e., two arbitrary data values $(t_i, x_i)$ and $(t_j, x_j)$ will have visibility, and consequently will become two connected nodes of the associated graph, if any other data $(t_q, x_q)$ placed between them satisfies:
\begin{equation}
x_q<x_j + (x_i-x_j)\frac{t_j-t_q}{t_j-t_i}
\label{vg}
\end{equation}
Note that the visibility graph is always connected by definition and also is invariant under affine transformations, due to the mapping method. On the other hand, ordered (periodic) and random series convert into regular and random exponential graphs, respectively: thus order and disorder structure in the time series seem to be inherited in the topology of the visibility graph. It is also shown that for a fractal time series, the distributions follow a power law form as $p_k \sim k^{-\gamma}$ \cite{Lacasa2009epl}, such that the Hurst exponent, $H$, of the series is linearly related to $\gamma$, i.e. $\gamma=5-2H$ for fGn and $\gamma=3-2H$ for fBm processes.

An alternative (and much simpler) algorithm is the horizontal visibility graph (HVG) \cite{Lacasa2009pre}, in which a connection can be established between two data points $i$ and $j$, if one can draw a \textit{horizontal} line in the time series joining them that does not intersect any intermediate data height, $x_q$,
by the following geometrical criterion:
\begin{equation}
x_i,x_j>x_q \text{   for all $q$ such that $t_i < t_q < t_j$}
\label{hvg}
\end{equation}
Because of the simplicity of the HVG, some features of the resulting graph has been calculated, analytically. It has been shown in \cite{Lacasa2009pre} that for an uncorrelated time series, the corresponding HVG is a small-world network \cite{Watts1998} with mean degree $\left\langle k\right\rangle=4$ and also its degree distribution, $p_k$, is as follows:
\begin{equation}
p_k \sim e^{-\lambda_c k}
\label{exp}
\end{equation}
with $\lambda_c=\ln(3/2)$ and these results are universal, i.e. independent of the probability distribution from which the series was generated. It has been also shown that the qualitative features of the HVG is the same as that of the VG, and quantitatively speaking, horizontal visibility graphs will have typically `less statistics' than the visibility graphs \cite{Lacasa2009pre}.

In this article, we will show that the visibility graph (VG) algorithm may not be a well-defined method to extract correlation information of a time series and its statistics is not essentially the same as that of the HVG. Accordingly, we will focus our attention into the horizontal visibility graph algorithm, and investigate correlation structure of various fractional processes. We will demonstrate that the HVG degree distributions of fractional processes have parabolic exponential forms. We also find that the corresponding fitting parameter can be used to estimate the value of the Hurst exponent. We further take into account other topological features, such as maximum eigenvalue of the adjacency matrix and the degree assortativity, and show that they all can be used to predict the Hurst exponent of a fractional time series, with an exception for anti-persistent fGn processes, in which no significant changes exist in the topological characteristics of the graph. Accordingly, we solve this problem by calculating the Spearman correlation coefficient between the nodes' degrees and their corresponding data values in the original time series.
\begin{figure}[t]
\begin{center}
\includegraphics[width=8cm]{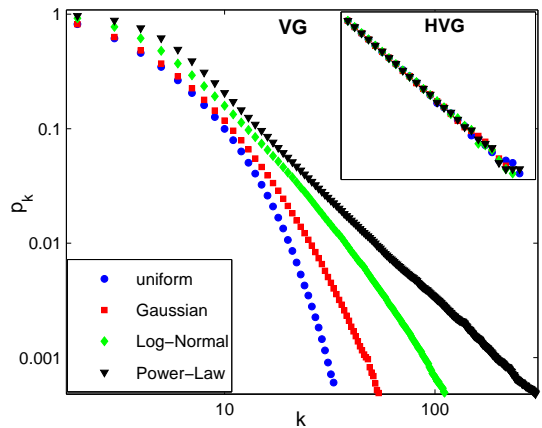}
\caption{The VG degree distributions, $p_k$, for uncorrelated (white) noises with different probability distributions: a uniform (circles), a normalized Gaussian (squares), a log-normal with parameters of $\mu=10$ and $\sigma=1$ (diamonds), and also a power-law of exponent $-3$ (triangles). The inset also shows the HVG degree distributions associated to such white noises, all with the same functional form of Eq.~\ref{exp}.}
\label{dist_VG}
\end{center}
\end{figure}

\section{Results and Discussions}%
In order to investigate the correlation information of fractional processes, we construct such time series with three different methods, a generic $1/f^{\delta}$ noise with Fourier filtering method (FFM) \cite{Feder1988}, a deterministic fBm process of Weierstrass-Mandelbrot function (WM) \cite{Berry1980}, and a stochastic fBm process with successive random addition method (SRA) \cite{Petigen1988}. We also use the detrended fluctuation analysis (DFA) exponent notation, $\alpha$, which is a generalization of the Hurst exponent. For a fGn, $\alpha\in(0,1)$, and the Hurst exponent, $H$, is equal to $\alpha$, and for a fBm, $\alpha\in(1,2)$, and we have $H=\alpha-1$ \cite{Peng1994}. In the followings, we apply the HVG algorithm to map fractional time series of size $N=10^6$. Also, all calculations have been performed by averaging over $50$ different realizations.

\subsection{Degree Distribution}
In this section, we first show that the VG algorithm strongly depends on the probability distributions of the original time series. In Fig.~\ref{dist_VG}, we plotted the calculated VG degree distributions, $p_k$, for four uncorrelated processes with different probability distribution functions: a uniform, a normalized Gaussian, a log-normal \cite{Crow1988} (with parameters $\mu=10$ and $\sigma=1$) as well as a power-law (of exponent $-3$) distribution. Clearly, we find different degree distributions, in spite of the same correlation structure (all data are white noises with zero correlation). Therefore, we can conclude that the VG algorithm can not be applied to correctly extract correlation information of a time series. On the other hand, in the inset of Fig.~\ref{dist_VG}, we plotted the HVG degree distributions of these four uncorrelated processes, all with the same functional form of Eq.~\ref{exp}. This also indicates that the qualitative differences between VG and HVG are crucial. Such findings are in contrast with recent studies \cite{Lacasa2009pre}. In what follows, we focus our attention into the topological characteristics of the horizontal visibility graph associated to fractional processes which are constructed by three different methods of FFM, WM, and SRA (mentioned above).

\begin{figure}[t]
\begin{center}
\includegraphics[width=8cm]{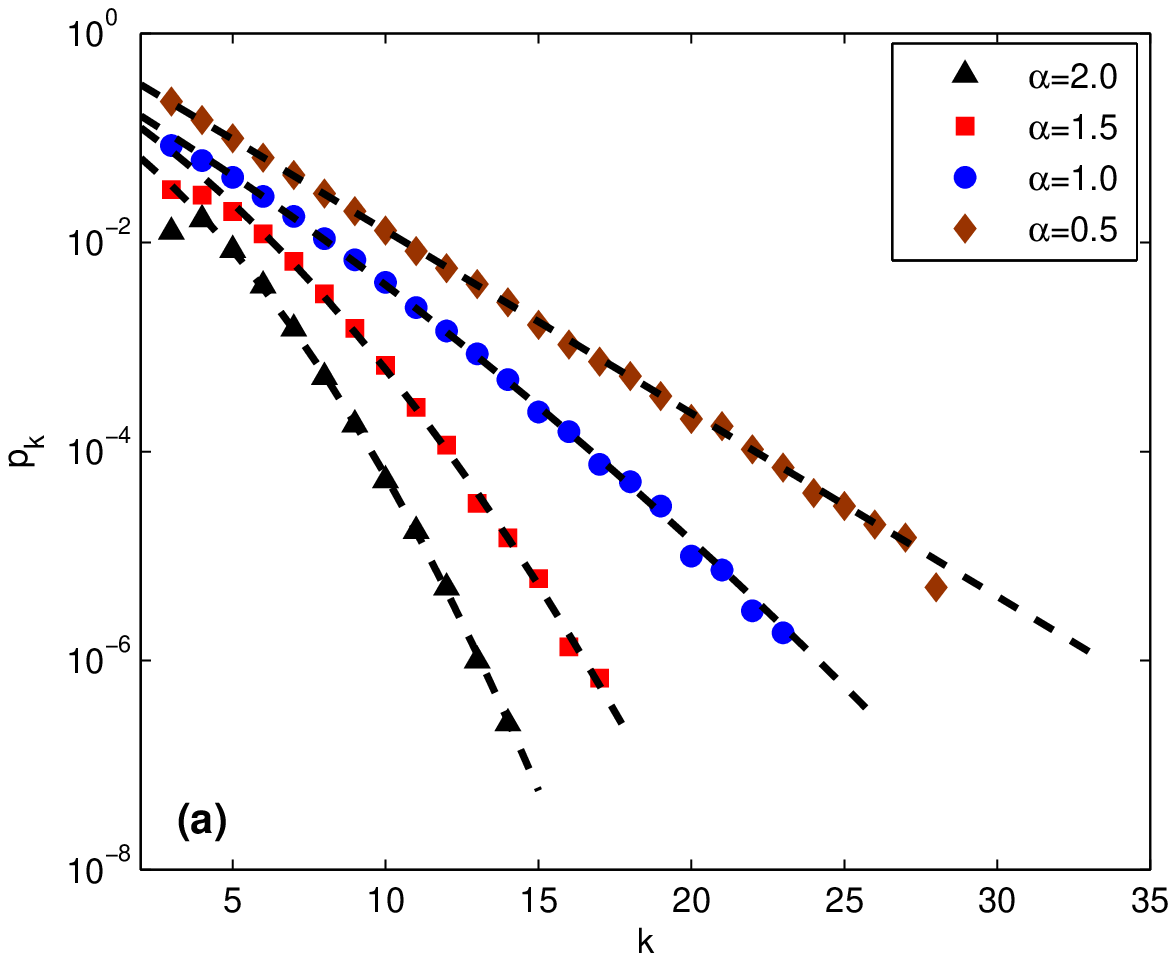}
\includegraphics[width=8cm]{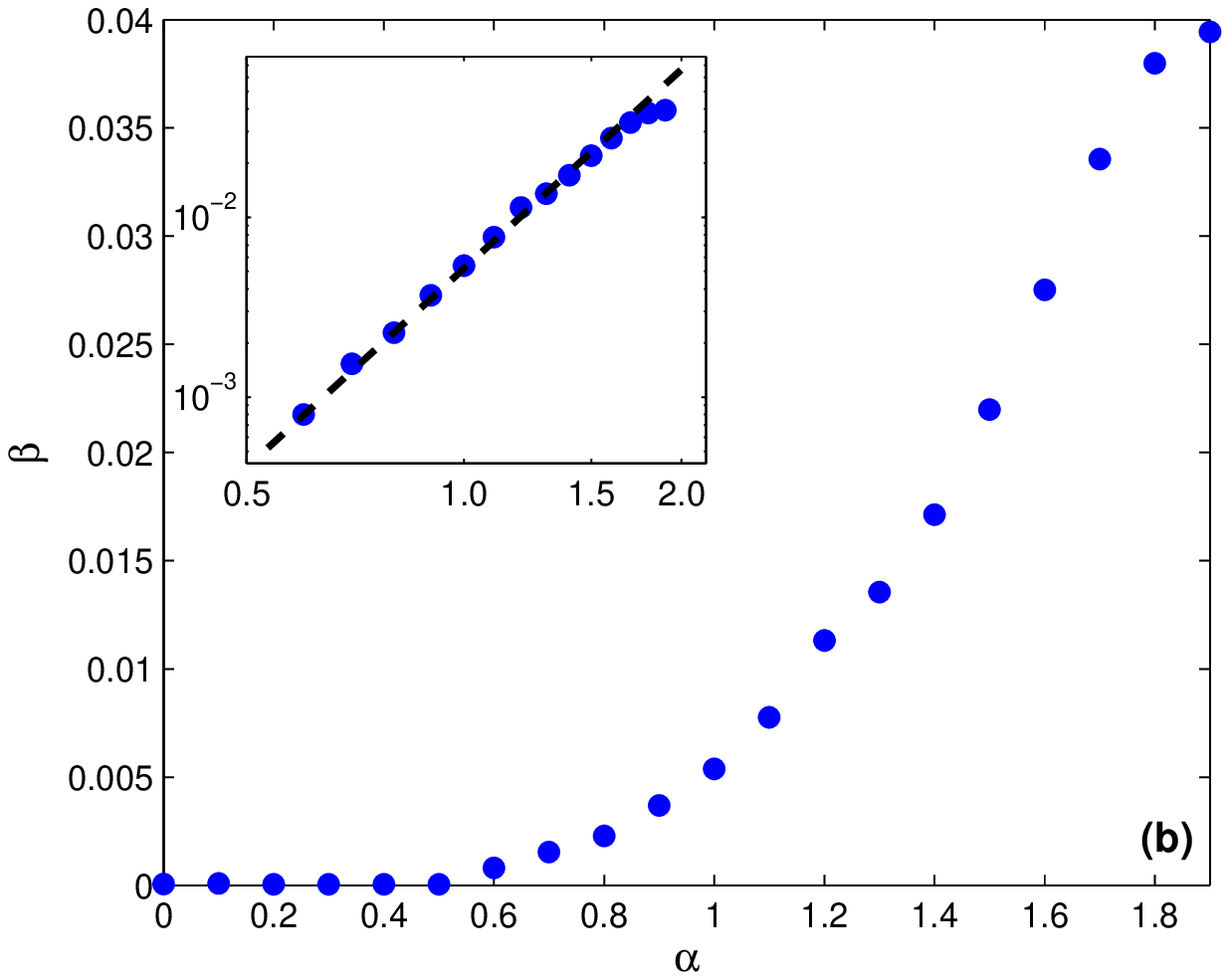}
\caption{(a) The HVG degree distributions associated to the fractional processes with $\alpha=0.5,1.0,1.5,2.0$, which are constructed by the FFM method. The dashed-lines show the parabolic exponentials of Eq.~\ref{PDF}. (b) The estimated fitting parameter, $\beta$, of Eq.~\ref{PDF}, for various $\alpha$. The inset shows the log-log plot of $\beta$, for various $\alpha>0.5$, and the dashed line also shows the estimated function of Eq.~\ref{beta_vs_alpha}. Clearly, the degree distributions are unable to discriminate between any anti-persistent fGns ($\alpha<0.5$)}.
\label{dist_HVG}
\end{center}
\end{figure}

In Fig.~\ref{dist_HVG}(a), we plotted HVG degree distributions of four fractional processes (using FFM) with different $\alpha=0.5,1.0,1.5,2.0$. The dashed lines represent the (proposed) parabolic exponential functions of the form
\begin{equation}
p_k\sim e^{-\lambda_ck}e^{-\beta k^2}
\label{PDF}
\end{equation}
where $\beta$ is the fitting parameter, and $\lambda_c$ is the same as in Eq.~\ref{exp}. We also plotted the fitting parameter, $\beta$, for various $\alpha$ in Fig.~\ref{dist_HVG}(b). We observe that for $\alpha \in(0,0.5)$, the distributions have nearly the same form with $\beta=0$, which corresponds to the exponential functions of Eq.~\ref{exp} for uncorrelated processes. Therefore, the HVG degree distributions are unable to discriminate between any anti-persistent fGn processes. On the other hand, $\beta$ increases with $\alpha$, which also shows that by increasing the correlation, the distributions decay faster. The log-log plot of the fitting parameter, $\beta$, is shown in the inset of Fig.~\ref{dist_HVG}(b), and the dashed line shows our suggested function for estimating $\alpha$, based on the value of $\beta$:
\begin{equation}
\beta\simeq 5\times10^{-3} \alpha^{3.5}
\label{beta_vs_alpha}
\end{equation}
where this estimated function works only for $\alpha>0.5$.

\subsection{Largest Eigenvalue}
\begin{figure}[t]
\begin{center}
\includegraphics[width=8cm]{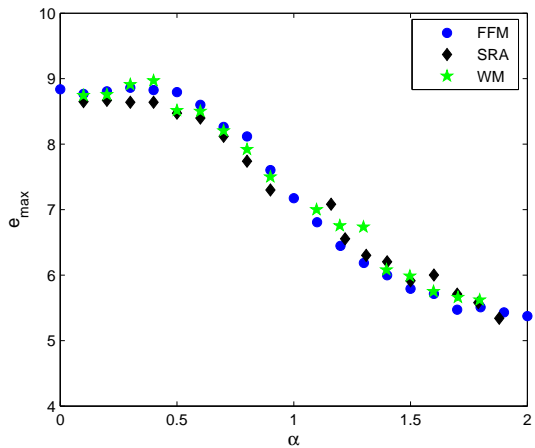}
\caption{The calculated maximum eigenvalue, $e_{max}$, of the adjacency matrix corresponds to the horizontal visibility graphs extracted from the fractional process versus the parameter $\alpha$. The different symbols represent different methods by which such fractional processes are constructed. Note that no significant changes can be detected for all anti-persistent fGns ($\alpha<0.5$).}
\label{eig_max}
\end{center}
\end{figure}

The largest eigenvalue, $e_{max}$, of the adjacency matrix (representing which nodes of a graph are adjacent to which other nodes) has emerged as a key quantity for the study of various topological characteristics of the complex networks \cite{Restrepo2005,May1972,Cvetkovic1990}. For networks with long-tail degree distributions, the largest eigenvalue is proportional to the maximum degree of the network \cite{Mihail2002,Chung2003,Dorogovtsev2003}. Here, the maximum eigenvalue of the adjacency matrices of the corresponding horizontal visibility graphs, extracted from three fractional processes with different methods of construction (FFM, SRA, and WM) are shown in the Fig.~\ref{eig_max}. The maximum eigenvalue is nearly constant for $\alpha\in(0,0.5)$ (anti-persistent fGn), and decreases with increasing $\alpha$, indicating that the presence of the correlation in the time series decreases the power of visibility of the data points in the original time series \cite{Lacasa2010}. Note that this topological property is also unable to discriminate between all anti-persistent fGn processes.

\subsection{Degree assortativity}
The assortativity coefficient, $r$, is the Pearson correlation coefficient \cite{Pearson1895} of the degrees at either ends of an edge and lies in the range $-1\leq r\leq1$ \cite{Newman2002}. This correlation function, $r$, is zero for no assortative mixing and positive (correlation between nodes of similar degree) or negative (correlation between nodes of different degree) for assortative or disassortative mixing, respectively. This correlation coefficient is defined as follows:
\begin{equation}
r=\frac{\Sigma_{jk}jk(e_{jk}-q_jq_k)}{\sigma^2_q}
\label{assort}
\end{equation}
where $e_{jk}$ is the fraction of edges that connect vertices of degrees $j$ and $k$, $q_k={(k+1)p_{k+1}}/{\left\langle k \right\rangle}$ is the distribution of the excess degree of the vertex at the end of an edge, and $\sigma_q$ is the standard deviation of the distribution $q_k$ \cite{Newman2002}.

\begin{figure}[t]
\begin{center}
\includegraphics[width=8cm]{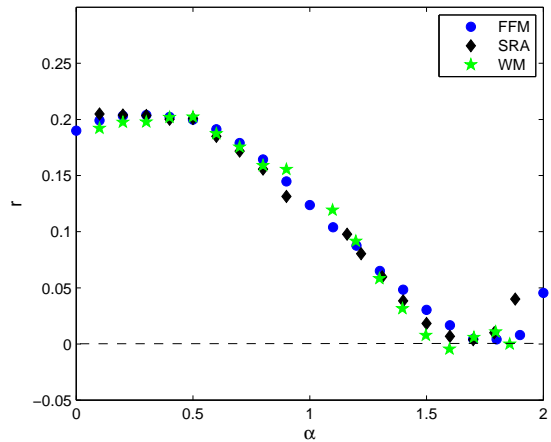}
\caption{The $\alpha$ dependency of the calculated degree assortativity, $r$, for the horizontal visibility graphs extracted from different fractional processes, constructed by three methods of FFM, SRA, and WM. Note that the correlation between the same degrees increases with decreasing $\alpha$.}
\label{assortativity}
\end{center}
\end{figure}

In Fig.~\ref{assortativity}, we plotted the assortativity coefficient for three fractional processes of FFM, SRA, and WM. As it can be seen, the assortativity is always positive and decreases with correlation. This means that the correlation between the same degrees decreases with increasing $\alpha$. For low $\alpha$'s, the HV graphs are highly assortative, thus, the hubs (nodes with highest degree) have better visibility on each other, which is called \textit{hub attraction}. The presence of hub attraction is due to the presence of more fluctuations in the original time series \cite{Lacasa2008}, which is a consistent result, here. For anti-persistent fGn, we find again a nearly constant value for $r$.

\subsection{Spearman correlation coefficient}
\begin{figure}[t]
\begin{center}
\includegraphics[width=8cm]{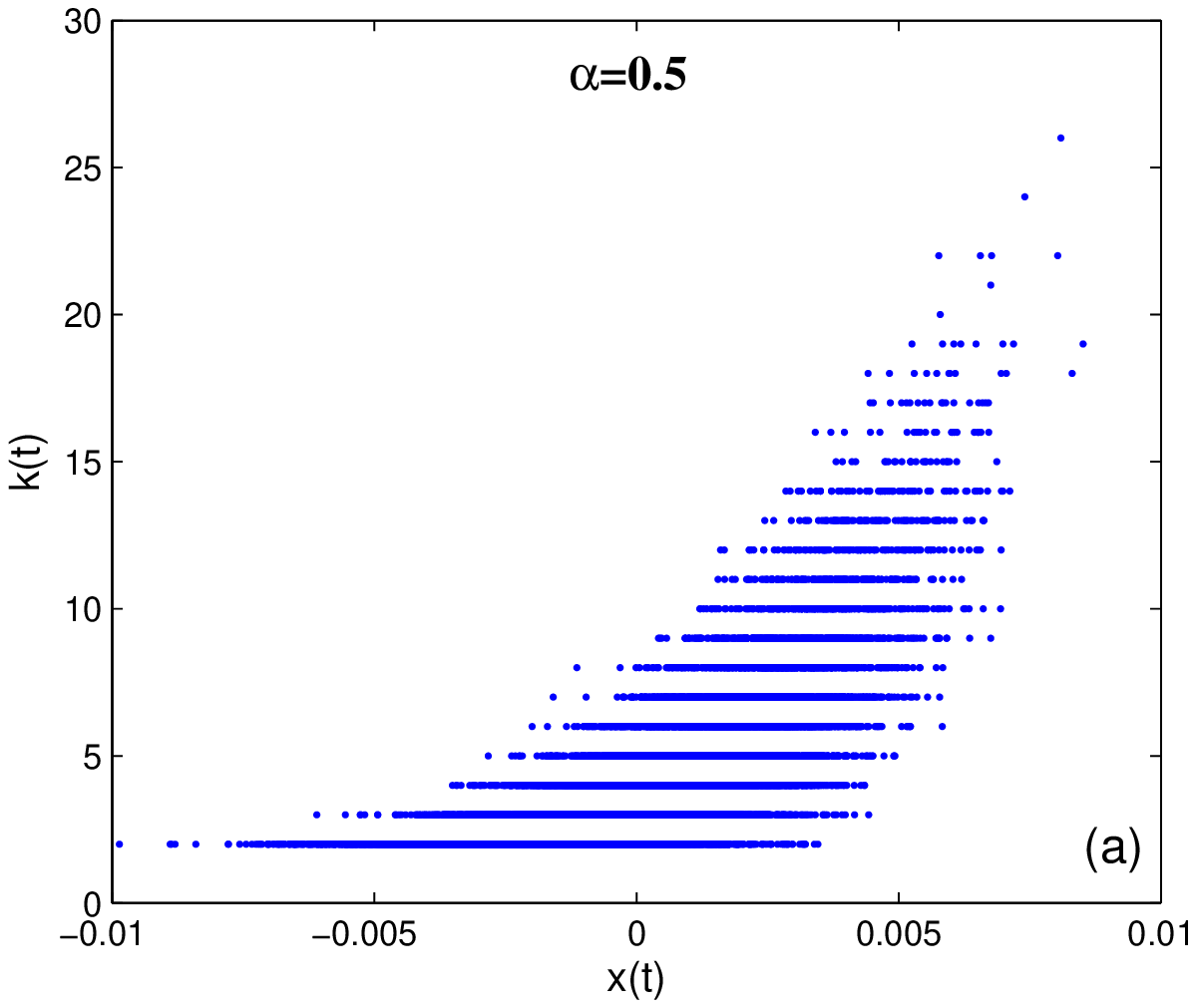}
\includegraphics[width=8cm]{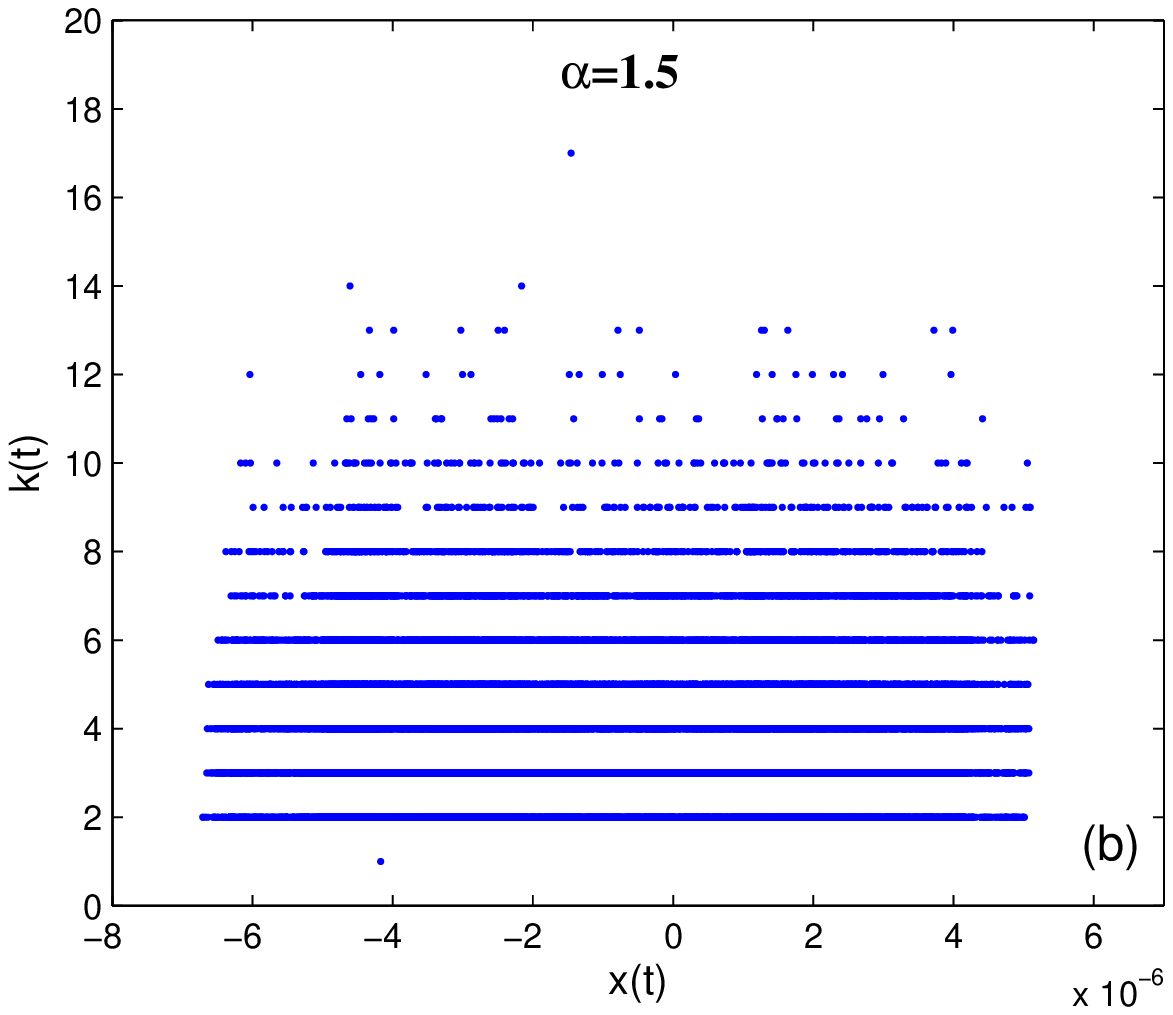}
\caption{The scatter plot of a FFM fractional time series, $x(t)$, and its corresponding HVG degree sequence, $k(t)$, for (a) $\alpha=0.5$ and (b) $\alpha=1.5$. Their Spearman's correlation coefficients are $0.77$ and $0.0$, respectively.}
\label{k_vs_x}
\end{center}
\end{figure}

\begin{figure}[t]
\begin{center}
\includegraphics[width=8cm]{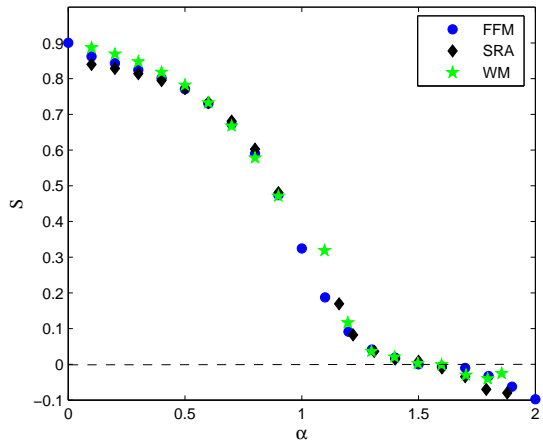}
\caption{The Spearman correlation coefficient, $S$, between fractional time series, $x(t)$, and the corresponding degree sequence, $k(t)$, for different $\alpha$, obtained from the horizontal visibility graph algorithm. The different symbols also represent different methods by which such fractional processes are constructed. We can see that this quantity is more sensitive to the correlation structure than other topological characteristics, especially for anti-persistent fGn region.}
\label{spear}
\end{center}
\end{figure}

Up to now, we have observed that no significant changes can be detected in all above-mentioned topological properties, for region $\alpha \in(0,0.5)$. Here, we try to find an appropriate quantity by which such processes can be discriminated. One of the most important properties of the visibility algorithm is that the time order of the original time series, $x(t)$, is preserved in the corresponding degree sequence, $k(t)$. Here, we search for the existence of any possible correlation between $x(t)$ and the corresponding $k(t)$. Fig~\ref{k_vs_x} shows the scatter plot of the degree sequence, $k(t)$, and the corresponding data point, $x(t)$ for $\alpha=0.5$ and $1.5$. To quantify this dependency, we calculate the Spearman correlation coefficient \cite{Spearman1904}, $S$, which is a statistical measure of the strength of a \textit{monotonic} relationship between paired data which lies in the range $-1\le S\le1$. It is worth to mention here that the Pearson correlation coefficient, which is a statistical measure of the strength of a \textit{linear} relationship between paired data, is not a good measure in this particular situation, because there is no necessity to have a linear relationship between $k(t)$ and $x(t)$. In Fig.~\ref{spear}, we plotted $S$ for different $\alpha$ for three fractional processes of methods FFM, SRA, and WM. We can see that this quantity is more sensitive to the correlation than the others, especially for anti-persistent fGn region.

In general, local maxima in the original time series typically have high degree, and local minima correspond to the low-degree nodes in the corresponding visibility graphs \cite{Lacasa2009pre}. In fractional processes, as $\alpha$ decreases, local maxima (minima) are more likely to contribute to the actual global maximum (minimum) values of the original series, i.e., the probability that high-value data points in the process have high degrees increases. Therefore, we can conclude that the correlation between values of the data points in the original time series and the degree of the corresponding nodes in the network positively increases, as $\alpha$ decreases. One can also see that the second derivative of the Spearman coefficient is approximately zero at $\alpha=1$, due to the sign change. At the end, we find that the Spearman correlation coefficient is a quantity which contains useful information about the correlation aspects of fractional time series.

\section{Conclusions}
In summary, we reinvestigated a recently proposed algorithm for mapping a time series into a (complex) network, called the visibility graph algorithm. We showed that the VG algorithm is strongly dependent on the distribution functions, of which the original time series are constructed. This result means that the VG algorithm is not an appropriate one to extract the correlation information of a process. On the other hand, by studying a much simpler algorithm, i.e. the horizontal visibility graph (HVG), we showed that the HVG and the VG are not the same neither statistically nor qualitatively. Afterwards, we focused our attention on the HVG algorithm, and studied some important topological properties of the fractional processes. We found that the HVG degree distributions extracted from such time series can well be fitted to parabolic exponential functions, with Hurst dependent fitting parameter. On the other hand, for all anti-persistent fractional Gaussian noises, $H \in(0,0.5)$, the same functional form is observed, indicating the inability of the HVG degree distributions to distinguish between such time series. By studying the maximum eigenvalues of the adjacency matrix, we observed that maximum degrees of the HVG decreases with increasing Hurst exponent. Further, we also calculated the degree assortativity of the HVG, and found that a positive correlation exists between nodes' degrees (hub attraction), and this correlation becomes weaker as $H$ increases. However, such quantities are also unable to discriminate between any anti-persistent fGn processes. Finally, we solved this issue by taking into account the correlation between degree of a node and its corresponding value in the original series. We applied the Spearman correlation coefficient to quantify such correlations, and observed that one can easily extract the correlation information of fractional Gaussian noises as well as fractional Brownian motions. It is noteworthy to mention here that the fractional processes used in this work were constructed by deterministic as well as stochastic methods. Finally, we can conclude that in spite of the inappropriateness of the VG algorithm, the HVG is a well-behaved formalism for extracting correlation information of stochastic time series.

\section*{Acknowledgments}
Support from Persian Gulf University Research Council is kindly acknowledged. I would also like to gratefully  acknowledge  the  many  useful  discussions  with  M. Ghaseminezhad Haghighi.


\begin{thebibliography}{50}

\bibitem{Kantz1997}
H. Kantz and T. Schreiber, \emph{Nonlinear Time Series Analysis} (Cambridge University Press, Cambridge, 1997).

\bibitem{Shumway2000}
R. H. Shumway and D. S. Stoffer, \emph{Time Series Analysis and Its Application} (Springer, New York, 2000).

\bibitem{Karlin1998}
S. Karlin and H. M. Taylor, \emph{An Introduction to Stochastic Modeling} (Academic Press, San Diego, 1998).

\bibitem{Lemons2002}
D. S. Lemons, \emph{An Introduction to Stochastic Processes in Physics} (Johns Hopkins University Press, Baltimore, 2002).

\bibitem{Kolmogorov1940}
A. N. Kolmogorov, C. R. (Doklady) Acad. URSS (N.S.) \textbf{26}, 115 (1940).

\bibitem{Mandelbrot1968}
B. B. Mandelbrot and J. M. Van Ness, SIAM Review \textbf{10}, 422 (1968).

\bibitem{Hurst1951}
H. E. Hurst, Trans. Amer. Soc. Civil Eng. \textbf{116}, 400 (1951).

\bibitem{Mandelbrot2002}
B. B. Mandelbrot, \emph{Gaussian Self-Affinity and Fractals} (Springer-Verlag, New York, 2002).

\bibitem{Embrechts2002}
P. Embrechts and M. Maejima, \emph{Self-similar Processes} (Princeton University Press, Princeton, 2002).

\bibitem{Karag2004}
T. Karagiannis, M. Molle, and M. Faloutsos, IEEE Internet Computing \textbf{8}, 57 (2004).

\bibitem{Karmeshu2004}
D. Karmeshu and A. Krishnamachari, "Sequence variability and long-range dependence in DNA: An information theoretic perspective" in \emph{Neural Information Processing, Lecture Notes in Computer Science}, (Springer, Berlin, 2004), pp. 1354–1361.

\bibitem{Robinson2003}
P. Robinson, ed., "Time Series with Long Memory" in \emph{Advanced Texts in Econometrics}, (Oxford University Press, 2003).

\bibitem{Varotsos2006}
C. Varotsos and D. Kirk-Davidoff, Atmospheric Chemstry and Physics \textbf{6}, 4093 (2006).

\bibitem{Alvarez2006}
E. Alvarez-Lacalle, B. Dorow, J.-P. Eckmann, and E. Moses, Proc. Nat. Acad. Sci. \textbf{103}, 7956 (2006).

\bibitem{Hurst1951}
H. E. Hurst, Trans. Amer. Soc. Civil Eng. \textbf{116}, 770 (1951); B. B. Mandelbrot and J. R. Wallis, Water Resour. Res. \textbf{5}, 242 (1969).

\bibitem{Holsch1988}
M. Holschneider, J. Stat. Phys. \textbf{50}, 963 (1988); J.-F. Muzy, E. Bacry, and A. Arn´eodo, Phys. Rev. Lett. \textbf{67}, 3515 (1991); J.-F. Muzy, E. Bacry, and A. Arn´eodo, Phys. Rev. E \textbf{47}, 875 (1993).

\bibitem{Peng1994}
C.-K. Peng, S. V. Buldyrev, S. Havlin, M. Simons, H. E. Stanley, and A. L. Goldberger, Phys. Rev. E \textbf{49}, 1685 (1994).

\bibitem{Vandewalle1998}
N. Vandewalle and M. Ausloos, Phys. Rev. E \textbf{58}, 6832 (1998); E. Alessio, A. Carbone, G. Castelli, and V. Frappietro, Eur. Phys. J. B \textbf{27}, 197 (2002).

\bibitem{Barabasi2002}
A.-L. Barab\'asi, \emph{LINKED: The New Science of Networks} (Perseus Publishing, Cambridge, Massachusetts, 2002).

\bibitem{Barrat2008}
A. Barrat, M. Barth\'elemy, and A. Vespignani, \emph{Dynamical Processes on Complex Networks} (Cambridge University Press, New York, 2008).

\bibitem{Pastor2007}
R. Pastor-Satorras and A. Vespignani, \emph{Evolution and Structure of the Internet} (Cambridge University Press, Cambridge, 2007).

\bibitem{Murray1993}
J. D. Murray, \emph{Mathematical Biology} (Springer Verlag, Berlin, 1993).

\bibitem{Montakhab2012}
A. Montakhab and P. Manshour, Europhys. Lett. \textbf{99}, 58002 (2012); Commun. Nonlinear Sci. Numer. Simul. \textbf{19}, 2414 (2014).

\bibitem{Zhang2006}
J. Zhang and M. Small, Phys. Rev. Lett. \textbf{96} 238701 (2006).

\bibitem{Lacasa2008}
L. Lacasa, B. Luque, F. Ballesteros, J. Luque, and J. Carlos Nu\~no, Proc. Natl. Acad. Sci. USA \textbf{105}, 4972 (2008).

\bibitem{Xu2008}
X. Xu, J. Zhang and M. Small, Proc. Natl. Acad. Sci. USA \textbf{105}, 19601 (2008).

\bibitem{Shirazi2009}
A. H. Shirazi, G. R. Jafari, J. Davoudi, J. Peinke, M. R. Rahimi Tabar, and M. Sahimi, J. Stat. Mech. \textbf{2009}, 07046 (2009).

\bibitem{marwan2009}
N. Marwan, J. F. Donges, Y. Zoua, R. V. Donner, and J. Kurths, Phys. Lett. A \textbf{373}, 4246 (2009).

\bibitem{Donner2010}
R. V. Donner, Y. Zou, J. F. Donges, N. Marwan, and J. Kurths, New J. Phys. \textbf{12}, 033025 (2010).

\bibitem{Donner2011}
R. V. Donner, M. Small, J. F. Donges, N. Marwan, Y. Zou, R. Xiang, and J. Kurths, Int. J. Bifurcation Chaos \textbf{21}, 1019 (2011).

\bibitem{Manshour2015}
P. Manshour, M. R. Rahimi Tabar, and J. Peinke, J. Stat. Mech. \textbf{2015}, 08031 (2015).

\bibitem{Lacasa2009epl}
L. Lacasa, B. Luque, J. Luque, and J. C. Nu\~no, EPL \textbf{86}, 30001 (2009).

\bibitem{Lacasa2009pre}
B. Luque, L. Lacasa, F. Ballesteros, and J. Luque, Phys. Rev. E \textbf{80}, 046103 (2009).

\bibitem{Watts1998}
D. J. Watts and S. H. Strogatz, Nature \textbf{393}, 440 (1998)

\bibitem{Lacasa2010}
L. Lacasa and R. Toral, Phys. Rev. E \textbf{82}, 036120 (2010).

\bibitem{Feder1988}
J. Feder, \emph{Fractals} (Plenum Press, New York, 1988); C.-K. Peng \emph{et al.}, Phys. Rev. A \textbf{44}, 2239 (1991);  H. A. Makse, S. Havlin, M. Schwartz, and H. E. Stanley, Phys. Rev. E \textbf{53}, 5445 (1996).

\bibitem{Berry1980}
M.V. Berry, Z. V. Lewis, Proc. R. Soc. Lond. A, \textbf{370}, 459-484 (1980).

\bibitem{Petigen1988}
H.-O. Peitgen, D. Saupe, M. F. Barnsley, Y. Fisher, and M. McGuire, \emph{The science of fractal images} (Springer-Verlag, New York etc., 1988).

\bibitem{Crow1988}
E. L. Crow and K. Shimizu, \emph{Lognormal Distributions: Theory and Applications} (M. Dekker, New York, 1988).

\bibitem{Restrepo2005}
J. G. Restrepo, E. Ott, and B. R. Hunt, Phys. Rev. E \textbf{71}, 036151 (2005); Phys. Rev. Lett. \textbf{96}, 254103 (2006).

\bibitem{May1972}
R. May, Nature (London) \textbf{238}, 413 (1972); J. Feng, V. K. Jirsa, and M. Ding, Chaos \textbf{16}, 015109 (2006)

\bibitem{Cvetkovic1990}
D. Cvetkovic and P. Rowlinson, Linear Multilinear Algebra \textbf{28}, 3 (1990).

\bibitem{Mihail2002}
M. Mihail and C. Papadimitriou, Lect. Notes Comput. Sci. \textbf{254}, 2483 (2002).

\bibitem{Chung2003}
F. Chung, L. Lu, and V. Vu, Proc. Natl. Acad. Sci. U.S.A. \textbf{100}, 6313 (2003).

\bibitem{Dorogovtsev2003}
S. N. Dorogovtsev, A. V. Goltsev, J. F. F. Mendes, and A. N. Samukhin, Phys. Rev. E \textbf{68}, 046109 (2003).

\bibitem{Pearson1895}
K. Pearson, Proc. Roy. Soc. London \textbf{58}, 240 (1895).

\bibitem{Newman2002}
M. E. J. Newman, Phys. Rev. Lett. \textbf{89}, 208701 (2002).

\bibitem{Spearman1904}
C. Spearman, Amer. J. Psychol. \textbf{15}, 72 (1904).

\end{thebibliography}
\end{document}